# AADNMR: A Simple Method for Rapid Identification of Bacterial/ Mycobacterial Infections in Antibiotic Treated Peritoneal Dialysis Effluent Samples for Diagnosis of Infectious Peritonitis

Anupam Guleria,[1,‡] Nitin K Bajpai,[2,‡] Atul Rawat,[1] C L Khetrapal,[1] Narayan Prasad,[2,*] Dinesh Kumar,[1*]

[1]Centre of Biomedical Research and [2]Department of Nephrology, SGPGIMS, Raibareli Road, Lucknow-226014, India

**Running Title:** NMR based Diagnosis of Infectious Peritonitis

[‡] **Both the authors have contributed equally:**

**\*Authors for Correspondence:**

**Dr. Dinesh Kumar**
(Assistant Professor)
Centre of Biomedical Research (CBMR),
Sanjay Gandhi Post-Graduate Institute of Medical Sciences Campus,
Raibareli Road, Lucknow-226014
Uttar Pradesh-226014, India
Mobile: +91-9044951791, +91-8953261506
Fax: +91-522-2668215
Email: dineshcbmr@gmail.com
Webpage: http://www.cbmr.res.in/dinesh.html

**Dr. Narayan Prasad**
(Additional Professor)
Department of Nephrology
Sanjay Gandhi Post-Graduate Institute of Medical Sciences Campus,
Raibareli Road, Lucknow-226014
Uttar Pradesh-226014, India
Mobile: +91-9415403140
Fax: +91-522-440973
Email: narayan.nephro@gmail.com

**ABBREVIATIONS:**

**NMR**, Nuclear Magnetic Resonance; **CPMG**, Carr–Purcell–Meiboom–Gill; **TOCSY**, Total Correlation Spectroscopy; **HSQC**, Heteronuclear Single Quantum Correlation; **PD**, Peritoneal Dialysis; **AADNMR**, Add Antibiotic to detect by NMR

# Graphical Abstract For Online:

**AADNMR: A Simple Method for Rapid Identification of Bacterial/ Mycobacterial Infections in Antibiotic Treated Peritoneal Dialysis Effluent Samples for Diagnosis of Infectious Peritonitis**

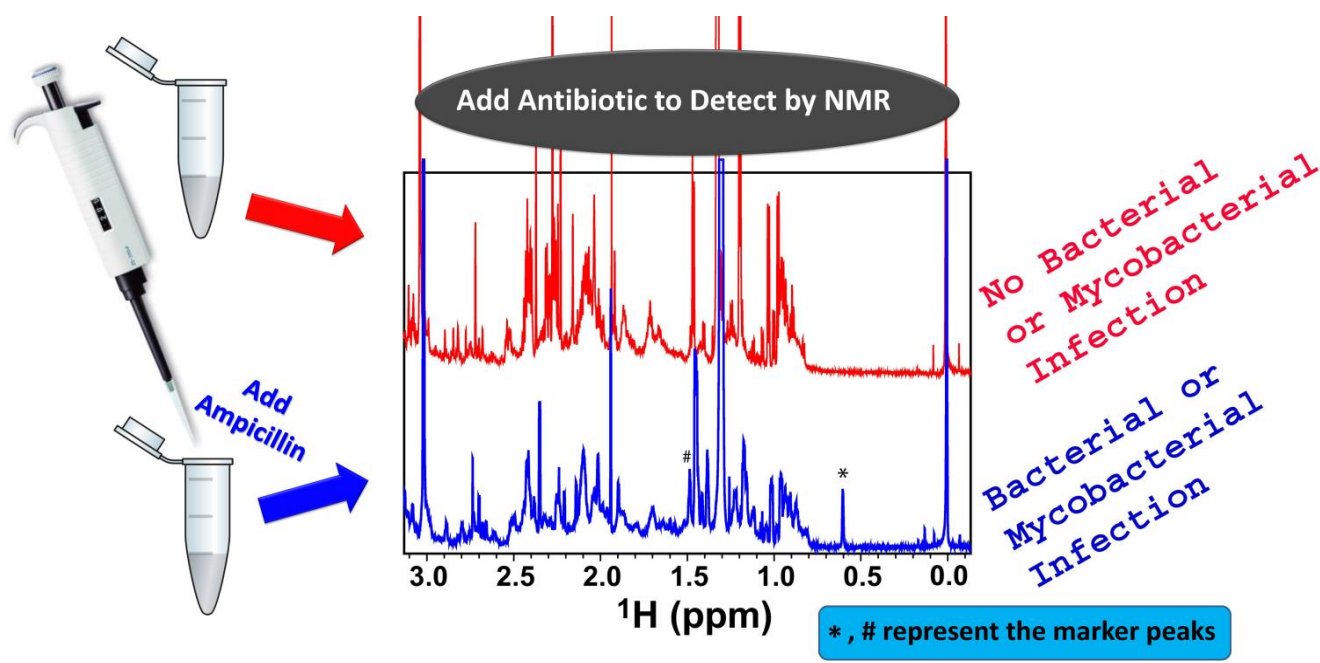

- Infectious peritonitis (bacterial/fungal) is the most common occurring complication associated with prolonged peritoneal dialysis (PD).
- PD effluent samples are generally obtained after the start of antibacterial treatment.
- Culture reports tend to be negative in the presence of intraperitoneal antibiotics.
- An efficient analytical tool is needed which can rapidly differentiate between bacterial and fungal peritonitis.
- In this context, a simple and efficient $^1$H NMR based method is reported –named here as "***add antibiotic to detect by NMR*** " or "**<AADNMR>**".

**Abstract:**

An efficient method is reported for rapid identification of bacterial/mycobacterial infection in a suspected clinical/biological sample. The method is based on the fact that the ring methylene protons of cyclic fatty-acids –constituting the cell membrane of several species of bacteria and mycobacteria- resonate specifically between -0.40 and 0.68 ppm region of the $^1$H NMR spectrum. These cyclic fatty acids are rarely found in the eukaryotic cell membranes. Therefore, the signals from cyclic ring moiety of these fatty acids can be used as markers (a) for the identification of bacterial and mycobacterial infections and (b) for differential diagnosis of bacterial and fungal infections. However, these microbial fatty acids when present inside the membrane are not easily detectable by NMR owing to their fast $T_2$ relaxation. Nonetheless, the problem can easily be circumvented if these fatty acids become suspended in solution. This has been achieved by abolishing the membrane integrity using broad spectrum antibiotics (including ampicillin). The suspended fatty acids are then detected by NMR to probe the infection. Therefore, the method has been given the name "**Add Antibiotic to Detect by NMR** or <**AADNMR**>". The method has been tested here using both Gram +ve and Gram -ve bacterial strains and finally the utility of method is demonstrated for discriminating bacterial and fungal infections to aid timely diagnosis of infectious peritonitis –a life threatening complication associated with prolonged peritoneal dialysis.

## Introduction:

Rapid detection of microbial pathogens in clinical samples is crucial for directing appropriate antimicrobial therapy and improving patient care and associated outcomes [1]. Every hour the appropriate treatment is delayed in patients in critical care has been shown to increase mortality by ~10-15 % [2-6]; therefore, earlier detection of the pathogenic microorganism has the potential to greatly benefit patient care. Infectious peritonitis is one of the most serious complications associated with prolonged peritoneal dialysis (PD) therapy [3,7] –a technique used for treating the patients with end-stage renal failure (ESRF). Severe and prolonged infectious peritonitis is the major cause of mortality in PD patients [8] or permanent malfunctioning of peritoneal membrane and switching to haemodialysis. Therefore, considerable attention has been paid on prevention and treatment of PD-related infections [3,7]. According to treatment procedure, before the patient is discharged from the hospital, he is given an inhospital training (~10-14 days) on the continuous ambulatory peritoneal dialysis (CAPD) and is instructed to put intra-peritoneal antibiotics (generally a combination against gram positive and gram negative bacteria) as soon he notices cloudy effluent with any of these symptoms like abdominal pain, vomiting and/or fever. The condition may also be associated with low ultra-filtration. However in case if the cloudy effluent does not clear in three days (after instilling initial antibiotics) along with subsiding of symptoms, the patient is instructed to report to the treating physician for further investigations (i.e. total and differential leucocyte count with bacterial and fungal culture of effluent). The initial treatments of bacterial and fungal peritonitis differ in that: in bacterial peritonitis, intravenous antibiotics are added in addition to intra-peritoneal and wait for response, whereas in fungal peritonitis initial treatment is to remove the CAPD catheter and patient is put on antifungals. The fungal peritonitis is more serious (than the bacterial peritonitis) and it leads to death of the patient in approximately 25% or more of the episodes [3]. Therefore, there is an urgent need for a rapid method to differentiate bacterial and fungal peritonitis and not to wait for 48-72 hours for culture reports to come as this may affect the prognosis of PD patient.

In this context, an efficient $^1$H NMR based method –named here as **"Add Antibiotic to Detect by NMR" or "AADNMR"**- has been proposed and utilized for rapid identification of bacterial/mycobacterial infection in PD effluent samples. The method exploits the inherent difference present in the fatty acid composition of microbial and eukaryotic cell membranes and provides unambiguous information about the bacterial/mycobacterial infection very rapidly. The method has been tested here using both gram +ve and gram -ve bacterial strains and finally the utility of method has been demonstrated for the diagnosis of infectious peritonitis –a serious complication associated with prolonged peritoneal dialysis. The method helps the physician in two ways: (a) if it confirms the presence of bacterial/mycobacterial infection, the PD patient is continued on broad-spectrum antibacterials and (b) if it rules out the possibility of having bacterial/mycobacterial infection –which then possibly indicates the presence of fungal infection- the patient is immediately referred for the antifungal treatment. Overall, the method has its great implication to start timely treatment of infected PD patients.

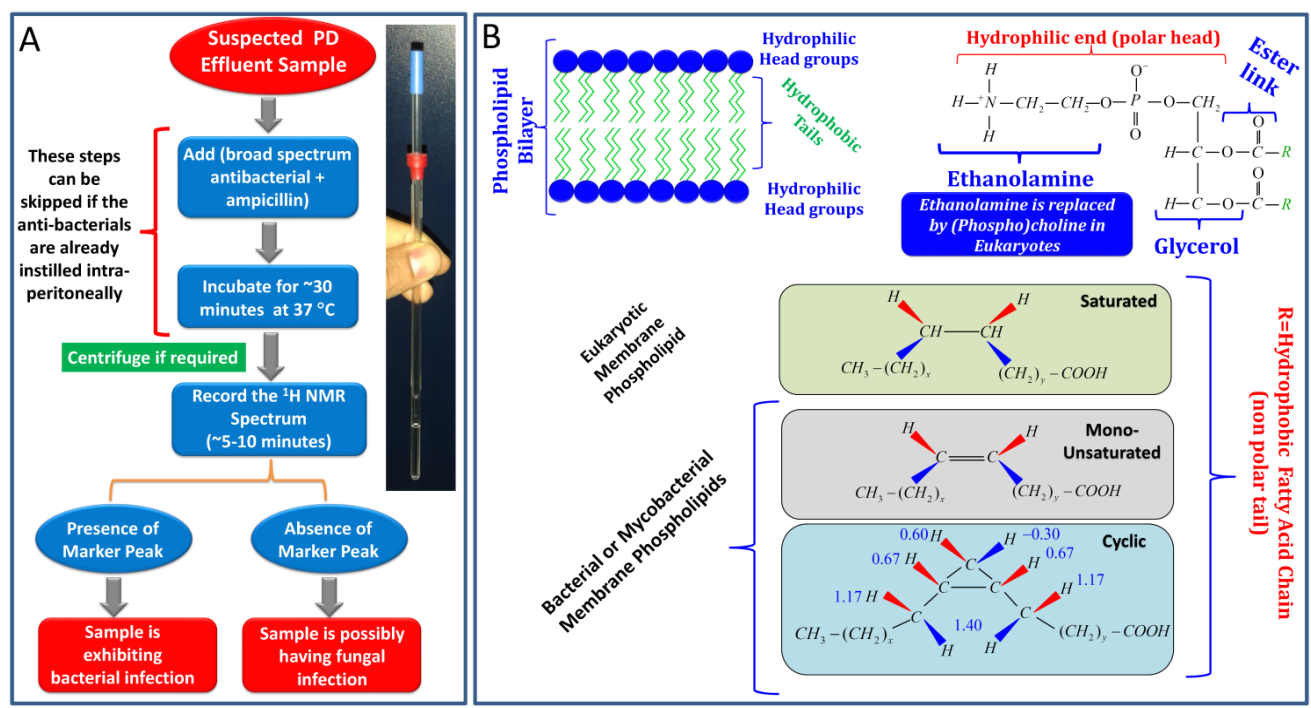

**Figure 1: (A)** Schematic showing the ¹H NMR based method –named here as **"add antibiotic to detect by NMR"** or **"AADNMR"**- for rapid identification of bacterial/mycobacterial infection in a biological/clinical sample. **(B)** Generalized structure of cell membrane with phospholipid bilayer (top right). The numbers in the right bottom panel (representing the cyclopropyl fatty acid geometry) represent the ¹H NMR chemical shifts of ring methylene protons as reported earlier [G. Knothe, Lipids (2006) 41, 393-396].

## Materials and Method:

**Chemicals:**

Deuterium oxide ($D_2O$) and sodium salt of trimethylsilylpropionic acid-$d_4$ (TSP) used for NMR spectroscopy were purchased from Sigma-Aldrich (Rhode Island, USA). Ampicillin –a broad spectrum antibacterial effective against both Gram +ve and Gram –ve bacteria– used here for *in vitro* NMR studies was also purchased from Sigma-Aldrich (Rhode Island, USA).

**Bacterial Cell Culture:**

The proposed <AADNMR> method was first tested on representative Gram-negative and Gram-positive bacterial strains named, respectively, *Escherichia coli* [ATCC 25922] and *Staphylococcus aureus* [ATCC 3160]. Both the strains were cultured in 25 mL of Luria-Bertani (LB) medium and were kept in a shaker incubator under identical conditions (37 °C, 200 rpm). Growth in these cultures were monitored by taking the absorbance (optical density, $OD_{600nm}$) on a Thermo Spectronic UV spectrophotometer at intervals of 1 hour. In each case, two aliquots of 1.0 mL culture were collected in the exponential phase (OD600 = 0.6); one aliquot was used as a control (representing live cell suspension) and the other aliquot was treated with ampicillin (50 µl of 10 mg/ml stock solution). Both the culture aliquots were again kept in the shaker incubator for about 2 hours under identical conditions (37 °C, 300 rpm). Finally, each culture aliquot was centrifuged at 12,000 rpm for 5 minutes to remove all cell debris and other contaminants. The supernatant part was decanted and stored at -20 °C until the $^1$H NMR experiments were performed.

**Clinical Samples:**

***Infected Urine Samples:*** To ensure its general utility, the method has first been validated on variety of infected urine samples obtained from UTI (Urinary Tract Infection) patients admitted in the Urology wards of Sanjay Gandhi Post Graduate Institute of Medical Sciences Lucknow. Inclusion criteria include urge to urinate frequently and need to urinate at night, painful burning sensation when urinating, discomfort or pressure in the lower abdomen, cloudy appearance and strong smell in urine, fever (typically lasting more than 2 days), impaired immune systems, or a history of relapsing or recurring UTIs, pain in the flank (pain that runs along the back at about waist level), vomiting and nausea. Each urine sample was divided into two parts of 1.0 mL each; one part was used as a control (representing live bacterial infection) and the other part was treated with ampicillin (50 µl of 10 mg/ml stock solution). Both the parts were incubated for about 1 hour under identical conditions (37 °C, 300 rpm). Finally, each part was centrifuged at 12,000 rpm for 5 minutes to remove all cell debris and other contaminants. The supernatant part was decanted and stored at -20 °C until the $^1$H NMR experiments were performed.

***Infected PD Effluent Samples:*** For demonstrating the clinical utility of the method, PD effluent samples (total 32 samples corresponding to 32 episodes) were obtained from 20 PD patients (n=20) admitted in the Nephrology wards of Sanjay Gandhi Post Graduate Institute of Medical Sciences, Lucknow. The study protocol was approved by the Hospital's Research Committee. All the selected PD patients were instilled with dialysate solution (Dianeal, 2.5 %) intraperitoneally containing dextrose. Among all the PD patients involved in this study (n=20), there was suspicion of having infectious peritonitis in 12 patients based on clinical symptoms and cloudy PD effluent (confirmed as per the guidelines of International Society of Peritoneal Dialysis) [3,7]. The PD patients with suspicion of having infectious peritonitis were given broad spectrum antibiotics instilled directly into the dialysate solution. Depending upon the criticality of infection, the antibiotics like Cefazolin, Tobramycin, Vancomycin, and ceftazidime were used for intraperitoneal instillation following standard treatment procedures. In each case, PD effluent sample was collected after a 4 hour dwell time, and was frozen and stored at a temperature of -20 °C, within 1-2 hours until the NMR measurements were performed.

**$^1$H NMR Spectroscopy:**

High Resolution NMR spectra were recorded at 298 K on a Bruker Avance III 800 MHz spectrometer (equipped with Cryoprobe). Standard relaxation edited 1D $^1$H NMR spectra were acquired using the Carr–Purcell–Meiboom–Gill (CPMG) pulse sequence [-recycle delay−π/2−(τ−π−τ)$_n$-acquisition] [9], with simple pre-saturation of the water peak, a total spin–spin relaxation time of 160 ms (n=400 and 2τ=400 μs), and a recycle delay (RD) of 5 sec. Each spectrum consisted of the accumulation of 64 scans and lasted for approximately 8 minutes. All the spectra were processed in Topspin-2.1 (Bruker NMR data Processing Software). Prior to Fourier Transformation (FT), the 1D $^1$H NMR data were zero-filled to 4096 data points and a sine–bell apodisation function was applied.

To confirm the assignment of marker peak as reported earlier [10], two-dimensional (2D) $^1$H-$^1$H total correlation spectroscopy (TOCSY) and $^1$H-$^{13}$C heteronuclear single quantum coherence (HSQC) spectra were acquired for some of the samples (including both bacterial cell cultures as well as PD effluents). Two-dimensional $^1$H–$^1$H TOCSY spectra were acquired in the phase sensitive mode using time proportional phase incrementation (TPPI), and the DIPSI2 pulse sequence for the spin lock [11]. 2048 data points with 16 transients per increment and 400 increments were acquired with a spectral width of 12 ppm in both dimensions. The RD between successive pulse cycles was 2 s and the mixing time of the DIPSI2 spin lock was 80 ms. The FIDs were weighted using a sine–bell-squared function in both dimensions and zero filled to 2048 and 1024 data points, respectively, in the $F_1$ and $F_2$ dimensions prior to FT. $^1$H–$^{13}$C HSQC spectra (phase sensitive with echo–antiecho) were recorded with inverse detection and $^{13}$C decoupling during acquisition [11]. A RD of 2.2 s was used between pulses and a refocusing delay equal to $1/(4*^1J_{C-H} = 1.75$ ms) was employed. 2048 data points with 64 scans per increment and 256 increments were acquired with spectral widths of 12 and 165 ppm in the $^1$H and $^{13}$C dimensions, respectively. The FIDs were weighted using a sine–bell-squared function in both dimensions and zero filled to 2048 and 1024 data points, respectively, in the $F_1$ and $F_2$ dimensions prior to FT.

**<AADNMR> Protocol:**

The **<AADNMR>** method for rapid identification of bacterial/mycobacterial infection has been illustrated in **Fig. 1A** and is based on the fact that several species of bacteria and mycobacteria contain cyclic fatty acids in their phospholipid bilayer membranes [12-18] which are rarely found in eukaryotic cell membranes except for some plant cells/organisms (http://lipidlibrary.aocs.org/Lipids/fa_cycl/index.htm). **Fig. 1B** displays the generalized features of various such phospholipid bilayer membranes. Various NMR based studies on structural analysis of cyclic fatty acids of microbial origin (including mycolic acids frequently found in mycobacterial cell membranes and contain di-substituted cyclopropane rings [10]) have shown that the ring methylene protons of the cyclic fatty acids resonate between -0.40 and 0.68 ppm region of the $^1$H NMR spectrum [19-21]. After an extensive analysis of $^1$H NMR chemical shifts of various small metabolites (including those in human, bacteria, and fungi metabolomics NMR databases) we found that except for long-chain compounds containing cycloalkane (C3 to C7) ring moiety, no small molecular weight compounds and metabolites resonate upfield to the 0.68 ppm region of the $^1$H NMR spectrum. Therefore the presence of a peak (peaks) between -0.40 and 0.68 ppm region of the $^1$H NMR spectrum may indicate identification of bacterial/mycobacterial infection. With this strategic aim, the present study was designed and signals from cyclic fatty acids were targeted. However when present within the membrane of a live bacterial/mycobacterial cell, these cyclic fatty acids are not easily detectable by NMR because of their very fast transverse relaxation (i.e. short $T_2$) owing to very long correlation time of the whole cell organism and therefore of associated membrane and its components. To produce signals from these fatty acids, the broad-spectrum antibiotics have been used, especially, the bactericidal antibiotics which disrupt the membrane (lipid bilayer) integrity of these bacterial cells (for details see these reference: [22,23]). The ampicillin induced bacterial cell membrane disruption brings these fatty acids into the sample solution for their facile detection by NMR. A point to be mentioned here that in the present study, we have found that the sample solution containing cyclic fatty acids of microbial origin (as produced here using antibiotic induced bacterial cell death) results into a strong $^1$H NMR signal resonating between 0.45 and 0.65 ppm. In the $^1$H-$^1$H TOCSY spectrum (see the Supplementary **Figure S1**), this peak was found to be correlated to another peak centred about 1.5 ppm (through vicinal trans coupling). In accordance with previous literature [10,19,20], the signal between 0.45 and 0.65 ppm represents cumulative NMR signal from trans methylene protons of cyclopropane ring moiety (as per the assignment of cyclopropane ring reported earlier [19] and also depicted in **Figure 1**). This particular peak has been referred here as the marker peak and has been labeled by asterisk (*) in subsequent Figures. The signal centred about 1.5 ppm has been attributed to vicinal *trans* protons attached to the carbons adjacent to the ring and has been labeled by symbol hash (#) in some of the Figures.

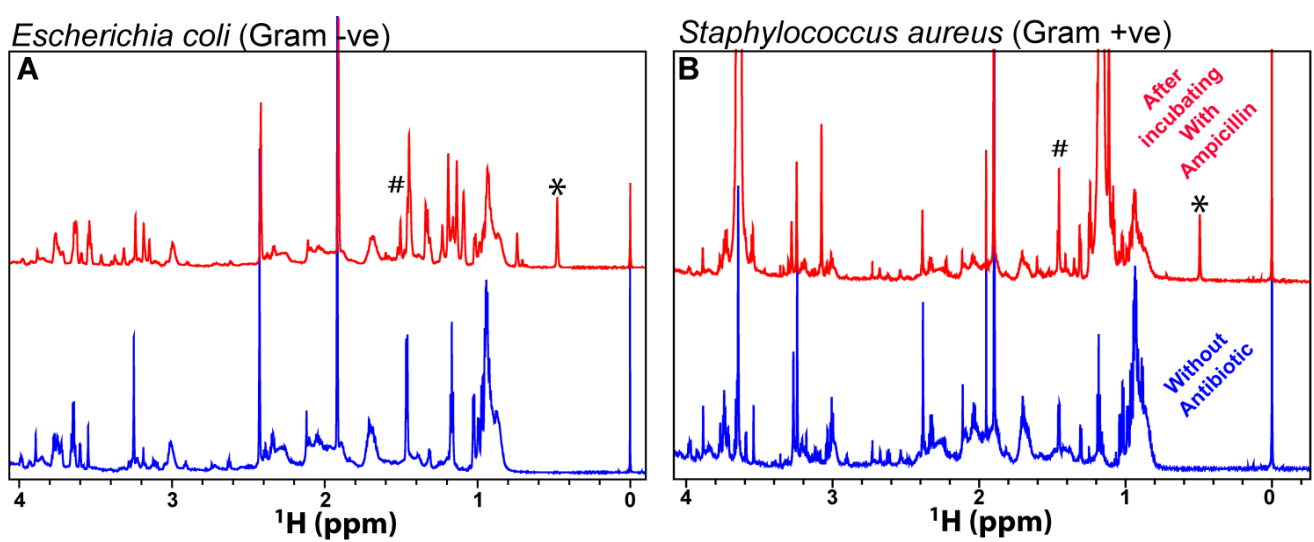

**Figure 2:** Representative one-dimensional (1D) $^1$H NMR spectra of extracellular metabolite solutions obtained, respectively, from ampicillin treated (red) and untreated (blue) bacterial cell cultures (grown in the LB media at 37 °C till the optical density of culture at 600 nm reaches to ~0.6): **(A)** *Escherichia coli* [ATCC-25922] (as a representative of Gram –ve bacteria) and **(B)** *Staphylococcus aureus* [ATCC-3160] (as a representative of Gram +ve bacteria).

Based on the facts and observations described above, the current **<AADNMR>** method has been designed. Briefly, the procedure includes adding a small amount of ampicillin (~ 10 μl of 100 mg/ml stock in 1 ml sample) to the suspected clinical/biological sample (if it is obtained without any medical treatment), incubating it for about 1 hour (at 37 °C and 200 rpm) and then recording a high resolution 1D $^1$H NMR spectrum. The spectrum is analysed simply through visual inspection and if the marker peak (as described above from 0.40 to 0.65 ppm region of the spectrum) is present, it simply confirms the presence of bacterial or mycobacterial infection. Typically a high resolution 1D $^1$H NMR spectrum can be acquired within a few minutes and since NMR requires minimal or no sample preparation or separation procedure, therefore for a clinical sample obtained after antimicrobial therapy has been started (like e.g. in the case of infectious peritonitis), the method can provide identification of microbial infection within about 20-30 minutes. The presence or absence of bacterial/mycobacterial infection further helps in differential diagnosis of bacterial and fungal infections and may aid physician treatment decisions. However the conventional culture based methods may produce false negative results in such situations.

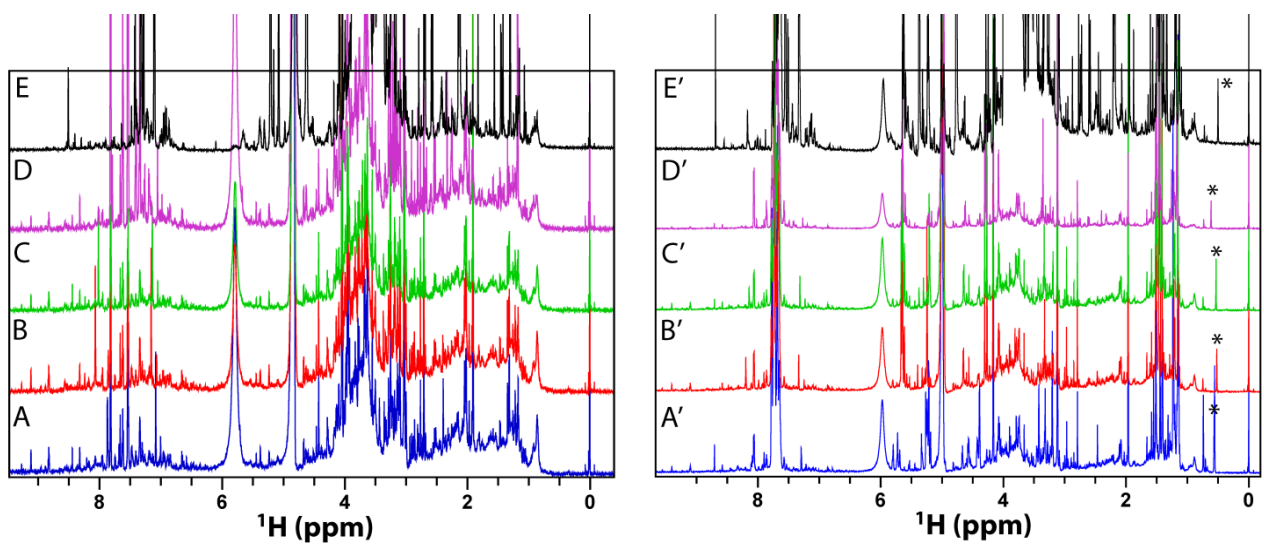

**Figure 3:** Stack plot of 1D $^1$H NMR CPMG spectra of urine samples of patients with gram (-ve) bacterial urinary tract infection (UTI) recorded at 800 MHz. **(A-E)** and **(A'-E')** represent, respectively, the control and ampicillin treated infected urine samples: **(A/A')** *Klebsiella pneumoniae*, **(B/B')** *Escherichia coli*, **(C/C')** *Pseudomonas aeruginosa*, **(D/D')** *Enterobacter aerogenes*, and **(E/E')** *Proteus mirabilis*.

## Results and Discussion:

**Validation of the Method through Experimental Testing:**

The proposed **<AADNMR>** method was first tested on representative Gram-negative and Gram-positive bacterial strains named, respectively, *Escherichia coli* and *Staphylococcus aureus* (also an important pathogen in catheter exit-site infection in peritoneal dialysis [24]). **Figure 2A** and **2B** displays their respective 1D ¹H NMR spectra before (blue) and after (red) treating with ampicillin. As evident from visual inspection, the marker peak (between 0.40 and 0.65 ppm region of the spectrum) is clearly visible in both the treated culture samples of Gram +ve and Gram –ve bacterial strains (prepared as described in the Materials and Method Section). To rule out the possibility that the observed marker peak is arising because of added antibiotic formulation, the 1D ¹H NMR spectra of antibiotics frequently used in this study (like Ampicillin, Vancomycin, Azolin etc) were also recorded. The spectra are shown in the Supplementary Material **Figure S2** and clearly show that the marker peak is not at all associated with any of these antibiotic formulations and is arising only because of the antibiotic induced bacterial cell lysis.

To further ensure that the marker peak is a reliable metabolic signature for identifying bacterial/mycobacterial infection, the method was further tested on urine samples obtained from Urinary tract infection (UTI) patients. **Figure 3** displays the 1D ¹H NMR spectra of infected urine samples. The spectra in **Fig. 3A-3E** represent the infected urine samples obtained without involving any antibiotic treatment, whereas the spectra in **Fig. 3A'-3E'** represent the infected urine samples treated *ex vivo* with ampicillin (a broad spectrum antibiotic active against both Gram +ve and Gram –ve bacteria). As evident, the marker peak is present in all the ampicillin treated infected urine samples, whereas, it is totally absent in untreated infected urine samples.

To ensure the microbial (bacterial or mycobacterial) origin of marker peak and its sole association with antibiotic (antibacterial and antimycobacterial) treatment, some additional experiments (including some positive and negative controls experiments) were carried out as well. The various experimental results and observations are described in **Appendix I** of the Supplementary material **(Figures S3-S9)**.

**Clinical Utility in the Diagnosis of Infectious Peritonitis:**

Infectious peritonitis is well-known cause of mortality in PD patients [8]. Therefore, timely diagnosis of PD related infections is crucial to save the patient's life via early intervention. According to the treatment procedure described in the introduction part, (as per the guidelines of ISPD i.e The International Society of Peritoneal Dialysis[3,7]), the PD effluent samples are mostly obtained after the antibiotic treatment has been started. Therefore the established methods [25] used for identification of infection may provide false negative results as they are all based on recovering and culturing of microorganisms from patient's body fluid. This prompted us to explore the ¹H NMR spectroscopy in the diagnosis of PD effluent. A simple ¹H NMR based method -as described in **Fig. 1-** has been proposed which enables the identification of bacterial/mycobacterial infection in a clinical or biological sample obtained after the antibiotic treatment therapy has been started as is the case with infectious peritonitis treatment procedure. The method –named here as **<AADNMR>**– can be used to rule in or rule out the presence of bacterial/mycobacterial infection in a PD effluent sample, therefore, would greatly aids the physician's treatment decision. The method was applied to analyse 32 PD effluent samples (corresponding to 15 episodes of 8 normal PD patients and 17 episodes of 12 suspected PD patients having infectious peritonitis). The various clinical details are tabulated below:

**Table 1.** Clinical details of PD effluent

| | PD effluent | | | |
|---|---|---|---|---|
| | **No infection** | **Bacterial infection** | | **Fungal infection** |
| | | **Without Antibiotic** | **With Antibiotic** | |
| Number of Episodes (No. of Patients) | 15 (n=8) | 3 (n=3) | 12 (n=7) | 2 (n=2) |
| Sample turbidity | No | Yes | Yes | Yes |
| Bacterial Culture Test | −ve | Gram +ve (Bacillus) | −ve | −ve |
| Fungal Culture Test | −ve | −ve | −ve | +ve |

For all the 15 episodes of normal PD patients and 3 episodes of infectious peritonitis (not given the antibiotic treatment) the marker peak was not observed suggesting that there is either no infection (as in the case with former 15 episodes) or PD patients with infection are not given the intra-peritoneal antibiotic treatment (as in the case with latter 3 episodes). For remaining 14 episodes of 9 PD patients with suspicion of having infectious peritonitis and are given intraperitoneal antibiotic treatment, the **marker peak** was observed in 12 episodes indicating the presence of intraperitoneal infection (bacterial/mycobacterial) in those corresponding PD patients. However for remaining two episodes with fair suspicion of infectious peritonitis, the negative <**AADNMR**> tests simply ruled out the possibility of bacterial/mycobacterial infection and hinted towards the possibility of having fungal infection. Later on clinical microbiology laboratory tests confirmed that these fairly suspected PD patients were having fungal infection. The presence of **marker peak** also correlated well with the onset and course of infection in a patient with 2 episodes of bacterial peritonitis and with response to therapy. A typical demonstration of various experimental results is displayed in **Figure 4**.

Compared to ampicillin treated bacterial cell cultures, where the marker peak was mostly centred about 0.5 ppm, in PD effluent samples the marker peak was slightly downfield shifted mostly centred about 0.63 ppm. These slight differences in the chemical shifts may be attributed to varying fatty acid composition of bacteria (particularly the position of the cyclopropane ring) which is markedly affected both quantitatively and qualitatively by the nature of the medium and by the conditions under which the culture is grown (more details on this phenomenon can be seen in these references [14,15,17,26]). To check this possibility, the bacterial (*Staphylococcus aureus*) cells were cultured directly into the PD dialysate solution (2.5 %) and treated with ampicillin. The experimental procedure and results are described in Supplementary material **(Appendix II, Fig. S10)**. As evident, the marker peak -detected in the ampicillin treated culture sample- is slightly upfield shifted towards 0.5 ppm **(Fig. S10C)** compared to that detected in infected PD effluent sample obtained after intraperitoneal antibiotic treatment **(Fig. 4C)** suggesting that the varying environmental/culture conditions are modulating the synthetic chemistry of microbial cyclic fatty acids.

**Future Directions:**

In this preliminary study based on small sample size, the use of antibiotic solution containing ampicillin has been shown for rapid detection of microbial infection in an infected solution sample. However, the method needs further validation on large sample size (including samples containing mycobacterial and fungal infections) for its reliable and routine application in clinical diagnosis, especially, in the context of differential diagnosis. Currently we have used ampicillin for most of the *ex vivo* experiments, therefore in future, the method will also be checked and validated for its use with variety of other antibiotics. In view of the criticality associated with fungal infection (as in the case of fungal peritonitis), we will also attempt to explore an alternative and unambiguous $^1$H NMR based method for rapid identification of fungal infection in solution samples.

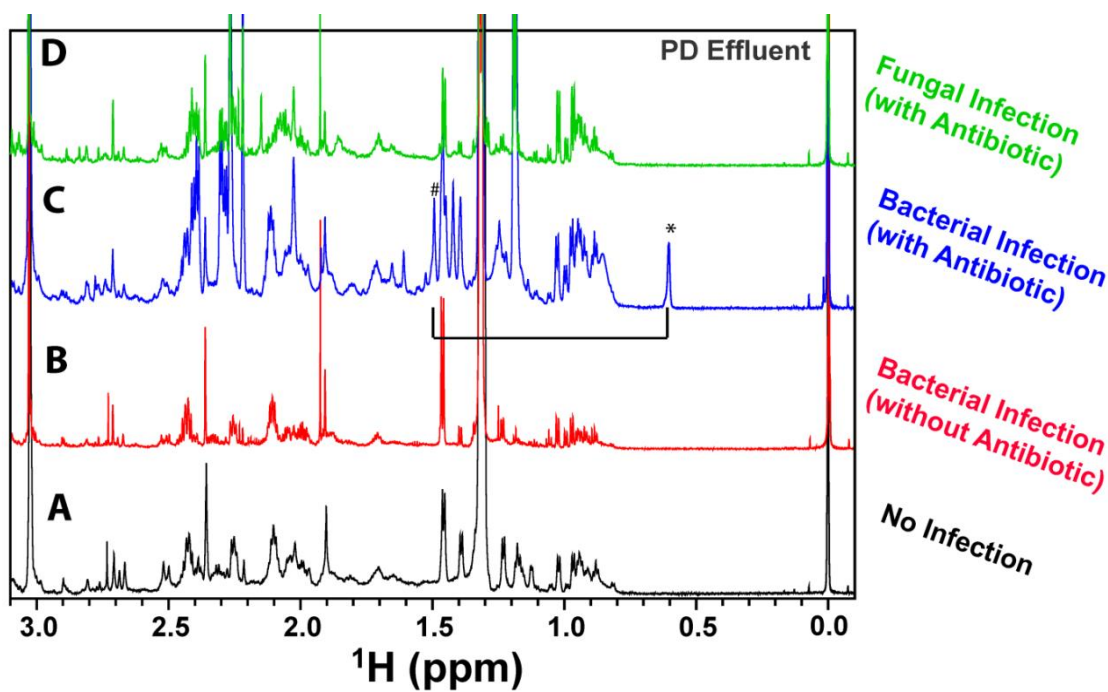

**Figure 4:** Stack plot of representative one-dimensional (1D) $^1$H NMR spectra of PD effluent (2.5 %, after 4 hour dwell time) from PD patients. **(A)** Represents the control PD effluent sample of a PD patient with no infection. **(B and C)** represent PD effluent samples from a PD patient having bacterial peritonitis and are obtained, respectively, before **(B)** and after **(C)** the intra-peritoneal antibiotic treatment has been started. In **(C)**, the presence of marker peak at ~0.61 ppm confirms the bacterial peritonitis which is otherwise difficult to predict through the routine cell culture based method if the patient has already been given the intraperitoneal antibiotic treatment. **(D)** Represents PD effluent sample from a PD patient having fungal peritonitis and given intraperitoneal antibiotic treatment. The absence of marker peak indirectly hints towards the fungal infection.

**Conclusion:**

In conclusion, an efficient and simple $^1$H NMR based method has been reported here for rapid identification of bacterial/mycobacterial infection in a suspected clinical/biological sample. The utility of the method has been demonstrated to analyse the PD effluent samples obtained after the antibiotic treatment has been started and the whole exercise –including sample preparation, data collection, and data analysis/interpretation (simply through visual inspection)- can be completed within about 20-30 minutes. The method aid the physician's treatment decision in two ways: (a) if it confirms the presence of bacterial/mycobacterial infection, the PD patient is continued on broad-spectrum antibacterials and (b) if it rules out the possibility of having bacterial/mycobacterial infection –which then possibly indicates the presence of fungal infection- the patient is immediately referred for the antifungal treatment. Further, the method can also be used for treatment monitoring and disease recovery in PD patients suffering from bacterial peritonitis. Overall, the method has its great implication to start timely treatment of infected PD patients to improve their surveillance.

The limitation of the method is that it is non-specific and does not provide any distinction between the infections caused by different bacterial strains. The other limitation of the method is that it may lead to false negative results when there is resistance in the bacterial strain against the antibiotic used for the cell lysis. However, the problem can be circumvented using a cocktail of broad spectrum antibiotics including ampicillin. Nevertheless we foresee that, the method because of its speed and simplicity will open up new avenues for a wide variety of clinical and biomedical applications as summarized below:

(a) The method can be used for treatment monitoring and recovery of infected patients in ICUs (like those with Urinary tract infection, respiratory tract infection, lung infection etc) where clinical samples (like urine, serum, sputum, nasal or pharyngeal swabs etc.) are obtained after the antibiotic treatment has been started.

(b) Because of its ability to rule in or rule out the presence of bacterial/mycobacterial infection in a clinical sample, the method can be used for differential diagnosis of bacterial meningitis and ventriculitis [38,39].

(c) The method can also be employed for rapid diagnosis of bloodstream infections by bacteria (known as bacteraemia) which is a serious medical problem associated with significant morbidity and mortality [40-44].

(d) The method can be used to test the antibacterial/bactericidal activity of a newly developed drug or to check if its mode of action is similar to ampicillin or other commonly used antibiotics. Likewise, the method can also be used to guide the treatment of patients in ICUs suffering from a bacterial/mycobacterial infection exhibiting multidrug-resistant (MDR). In such cases, the presence of marker peak in the affected body fluid sample (like blood, serum, urine, CSF, etc.) obtained after the medical treatment is started will help to confirm if the antibiotic combination is effective or not.

## Acknowledgement:

We acknowledge the Department of Science and Technology (DST), India for providing funds for the 800 MHz NMR spectrometer at Centre of Biomedical Research (CBMR), Lucknow, India. Dinesh Kumar would also like to acknowledge DST, India for providing him the research grant under SERC Fast Track Scheme (Registration Number: **SR/FT/LS-114/2011**).

# Supplementary Material:

## AADNMR: A Simple Method for Rapid Identification of Bacterial/ Mycobacterial Infections in Antibiotic Treated Peritoneal Dialysis Effluent Samples for Diagnosis of Infectious Peritonitis

Anupam Guleria,[1,‡] Nitin K Bajpai,[2,‡] Atul Rawat,[1] C L Khetrapal,[1] Narayan Prasad,[2,*] Dinesh Kumar,[1,*]

*[1]Centre of Biomedical Research and [2]Department of Nephrology, SGPGIMS, Raibareli Road, Lucknow-226014, India*

**Running Title:** Diagnosis of Infectious Peritonitis by NMR

[‡] **Both the authors have contributed equally:**

**\*Authors for Correspondence:**

**Dr. Dinesh Kumar**
(Assistant Professor)
Centre of Biomedical Research (CBMR),
Sanjay Gandhi Post-Graduate Institute of Medical Sciences Campus,
Raibareli Road, Lucknow-226014
Uttar Pradesh-226014, India
Mobile: +91-9044951791, +91-8953261506
Fax: +91-522-2668215
Email: dineshcbmr@gmail.com
Webpage: http://www.cbmr.res.in/dinesh.html

**Dr. Narayan Prasad**
(Additional Professor)
Department of Nephrology
Sanjay Gandhi Post-Graduate Institute of Medical Sciences Campus,
Raibareli Road, Lucknow-226014
Uttar Pradesh-226014, India
Mobile: +91-9415403140
Fax: +91-522-440973
Email: narayan.nephro@gmail.com

**Figure S1:** An overlay of the chemical shift matched regions of $^1$H-$^1$H TOCSY spectra of PD effluent samples obtained after 4 hours dwell time from a PD patient having bacterial infection. The blue and red spectra correspond to two episodes of the same patient where the PD effluent samples have been obtained with and without intraperitoneal antibiotic treatment. As evident from the presence of marker peak, the PD patient has developed intra-peritoneal bacterial/mycobacterial infection. The correlation peak between marker peak (*) and peak at 1.5 ppm (#) confirms the presence of cyclic fatty acids of microbial origin in accordance with the previous reports (see the text).

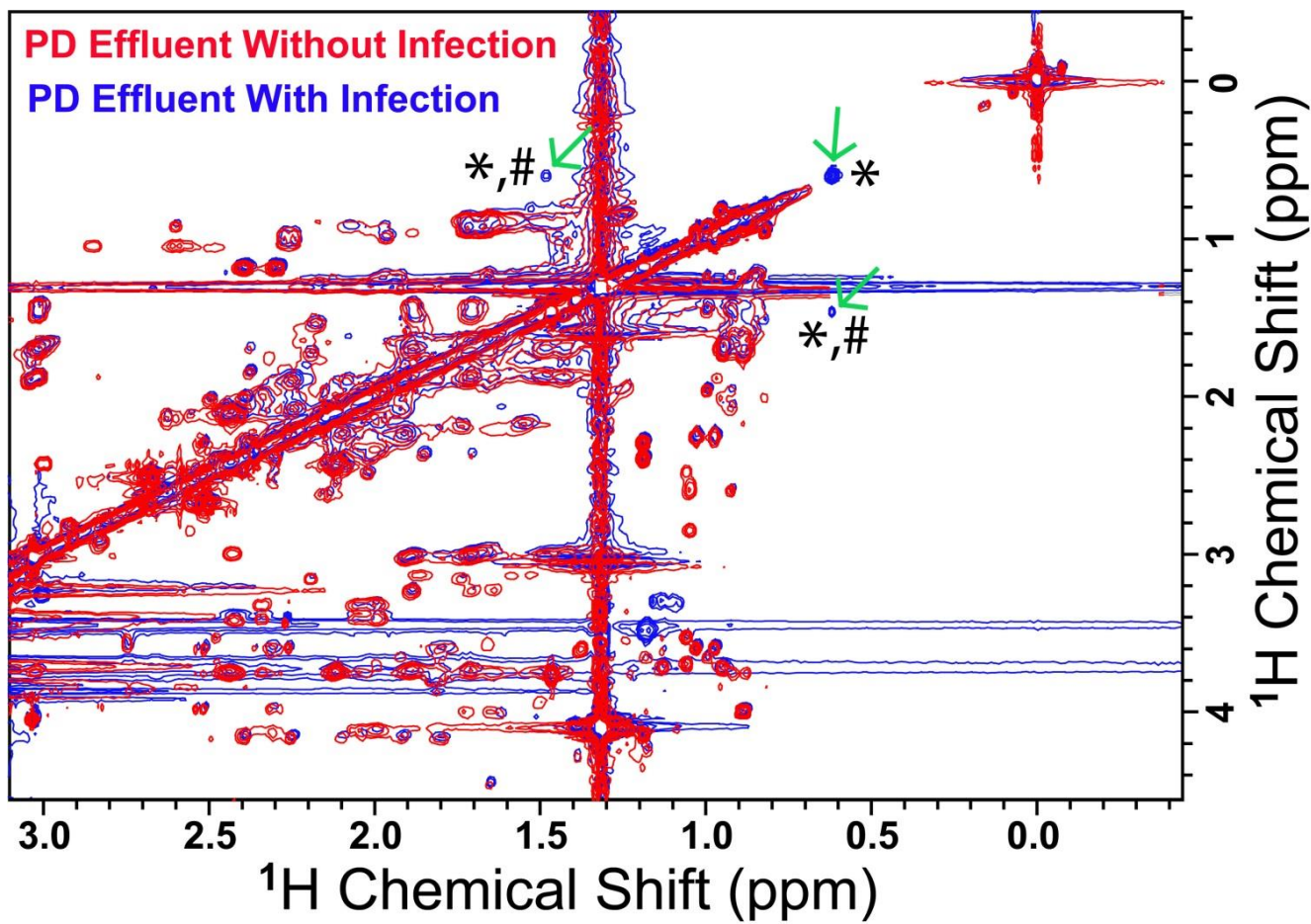

**Figure S2:** Stack plot of one-dimensional (1D) ¹H NMR spectra of some of the antibiotics (antibacterials) frequently used in the present study: **(A)** Ampicillin (effective against both Gram +ve and Gram –ve bacteria), **(B)** Azolin (from Biochem Pharmaceutical Industries Ltd, Mumbai, effective against Gram +ve bacteria) and **(C)** Vancomycin (from AstraZeneca, effective against Gram +ve bacteria). Spectra clearly reveal that there is no peak in the spectral region defined for the marker peak (i.e. between 0.40 and 0.68 ppm).

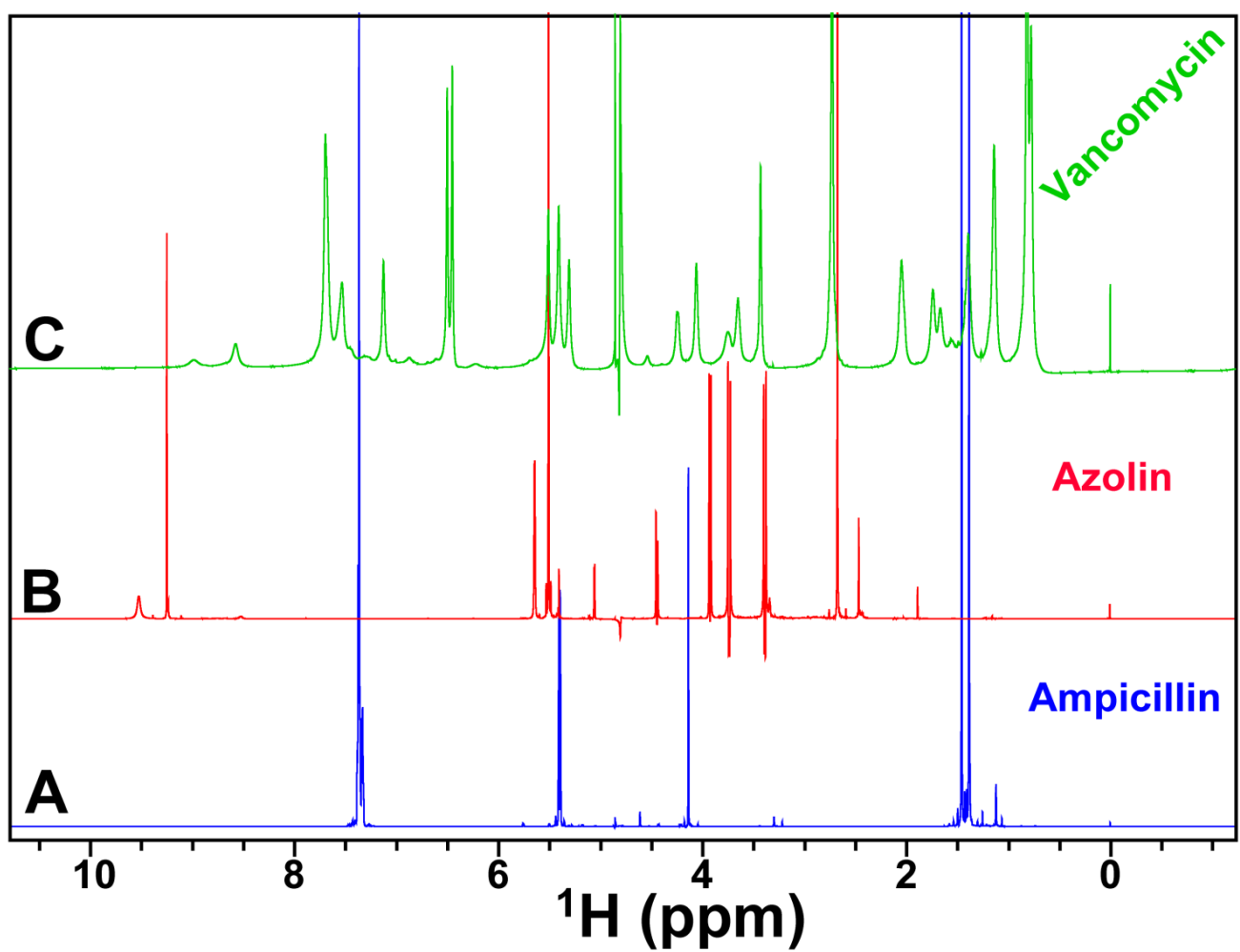

## *Appendix I:*

To ensure the microbial (i.e. bacterial or mycobacterial) origin of marker peak and its association with antibiotic treatment, some additional experiments were also carried out as summarized below:

1. To rule out the possibility that eukaryotic cells when treated with antibiotics may produce the marker peak (because of some kind of biochemical reaction), human embryonic kidney cells (HEK-293) were cultured in Dulbecco's Modified Eagle's Medium (DMEM, Invitrogen) containing 10 % foetal bovine serum (FBS, Invitrogen) and 1% Penicillin (antibacterial) and Streptomycin (antimycobacterial). 1 x $10^6$ cells were seeded in a T75 flask, when the cells reached 80% confluency, they were divided into two equal volumes: the first volume was directly used to record the 1D $^1$H CPMG NMR spectrum shown in **Fig S3A**, whereas the second volume was sonicated and centrifuged at 4 °C and 9000 rpm for 10 minutes, supernatant was used to record the 1D $^1$H CPMG NMR spectrum shown in **Fig S3B**. As evident, the marker peak is absent in both the conditions suggesting that the marker peak is not resulting from any biochemical reaction between the antibiotic and the eukaryotic cell membrane.

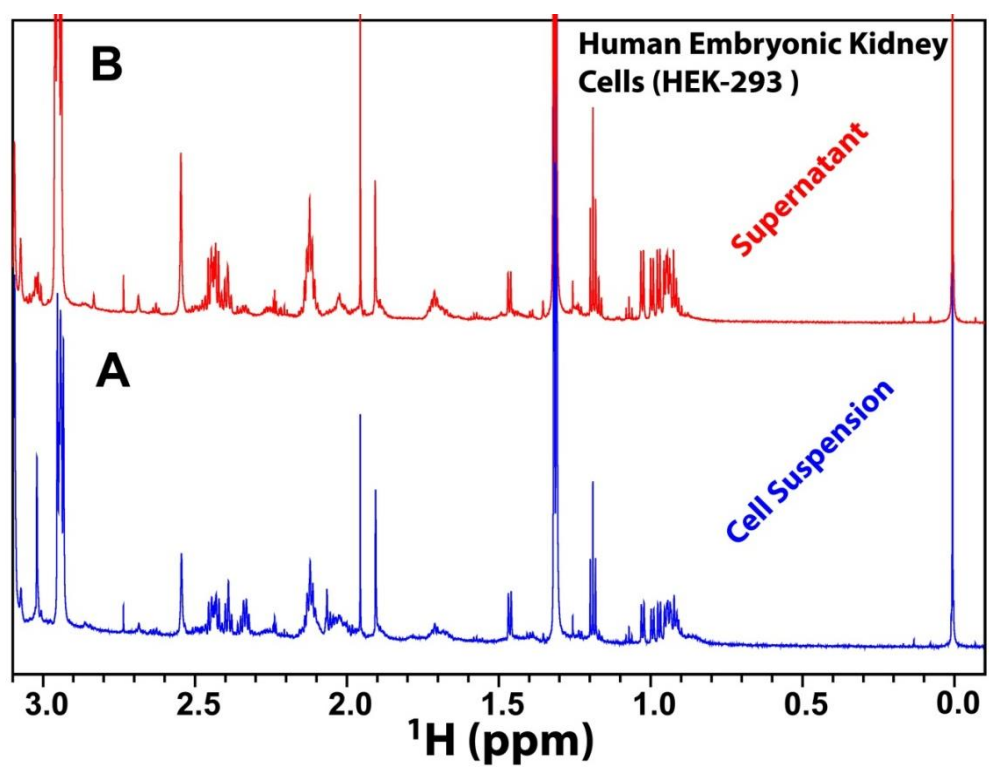

**Figure S3:** Stack plot of one-dimensional (1D) $^1$H NMR spectra of human embryonic kidney (HEK-293) cell line culture (grown in the presence of broad spectrum of anti-bacterials as described above): **(A)** cell suspension and **(B)** supernatant obtained after sonication (at 4 °C) and centrifugation at 9000 rpm for 10 minutes.

2. To rule out the possibility that a clinical/biological sample lacking bacterial/mycobacterial infection may produce the marker peak when treated with antibacterial, clean PD effluent sample was used to perform the **<u>negative control experiment</u>**. The clean PD effluent sample was transferred to two separate 2 ml centrifuge tubes (1 ml in each tube). One aliquot was directly used to record the 1D ¹H CPMG NMR spectrum shown in **Fig S4A**, whereas the second aliquot was treated with ampicillin (antibacterial) for 1 hour at 37 °C and used to record the 1D ¹H CPMG NMR spectrum shown in **Fig S4B**. As evident, the PD effluent sample treated with ampicillin does not produce the marker peak **(Fig S4B)** inferring that the marker peak is associated with bacterial/mycobacterial infection as stated in this paper.

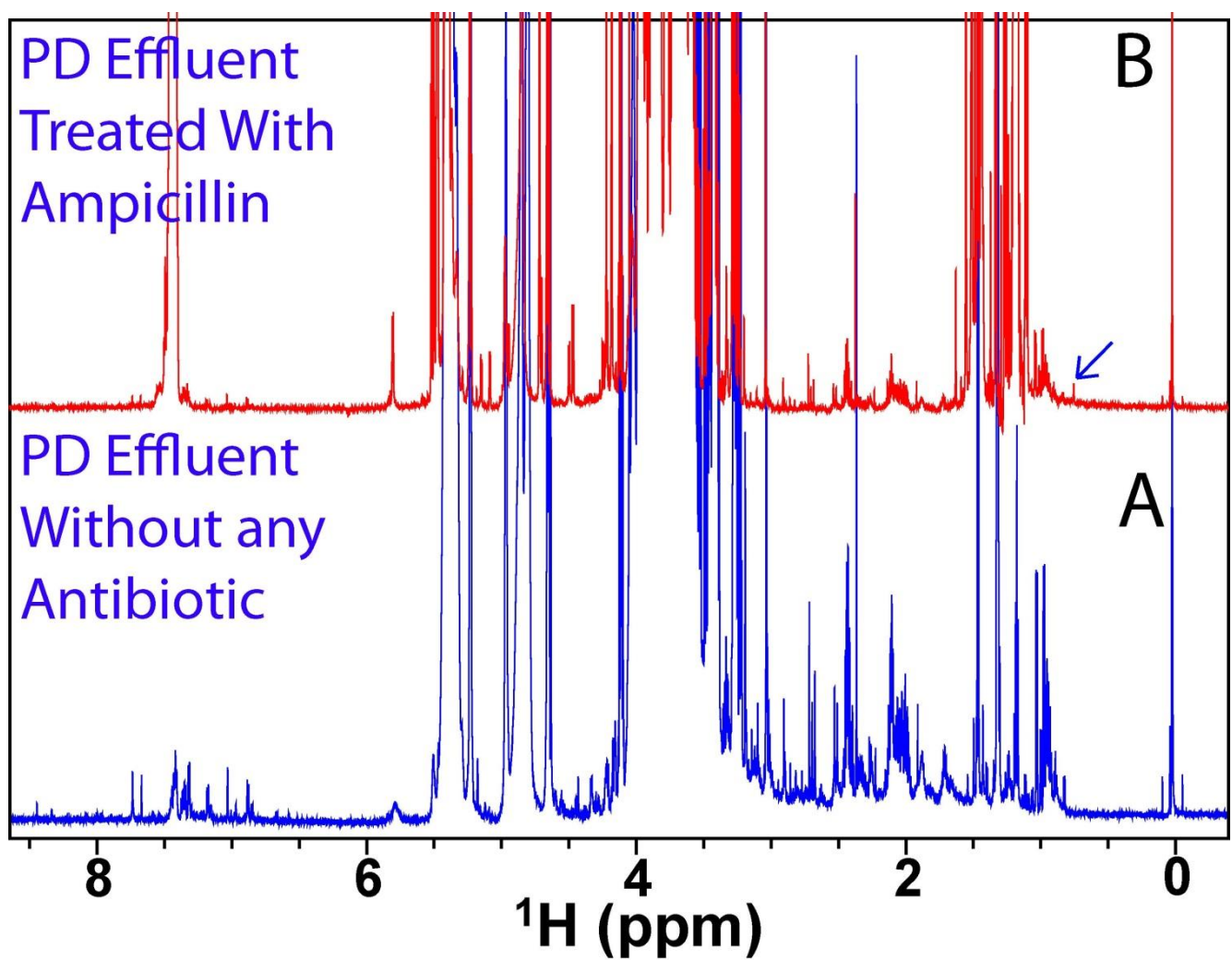

**Figure S4:** Stack plot of one-dimensional (1D) ¹H CPMG NMR spectra of fungal cell culture: **(A)** without ampicillin treatment and, **(B)** after incubating with ampicillin for 1 hour at 37 °C. The peak at ~0.73 ppm pointed here using a blue arrow is associated with ampicillin and does not belong to any microbial infection.

3. To rule out the possibility that a clinical/biological sample having fungal infection may produce the marker peak when treated with antibacterial, PD effluent sample having fungal infection was used to perform the **negative control experiment**. The fungal infected PD effluent sample was transferred to two separate 2 ml centrifuge tubes (1 ml in each tube). One aliquot was directly used to record the 1D ¹H CPMG NMR spectrum shown in **Fig S5A**, whereas the second aliquot was treated with ampicillin (antibacterial) for 1 hour at 37 °C and used to record the 1D ¹H CPMG NMR spectrum shown in **Fig**

**S5B**. As evident, the PD effluent sample containing fungal infection treated with ampicillin does not produce the marker peak **(Fig S5B)**. This clearly indicates the microbial (bacterial/mycobacterial) origin of the marker peak as stated in this paper.

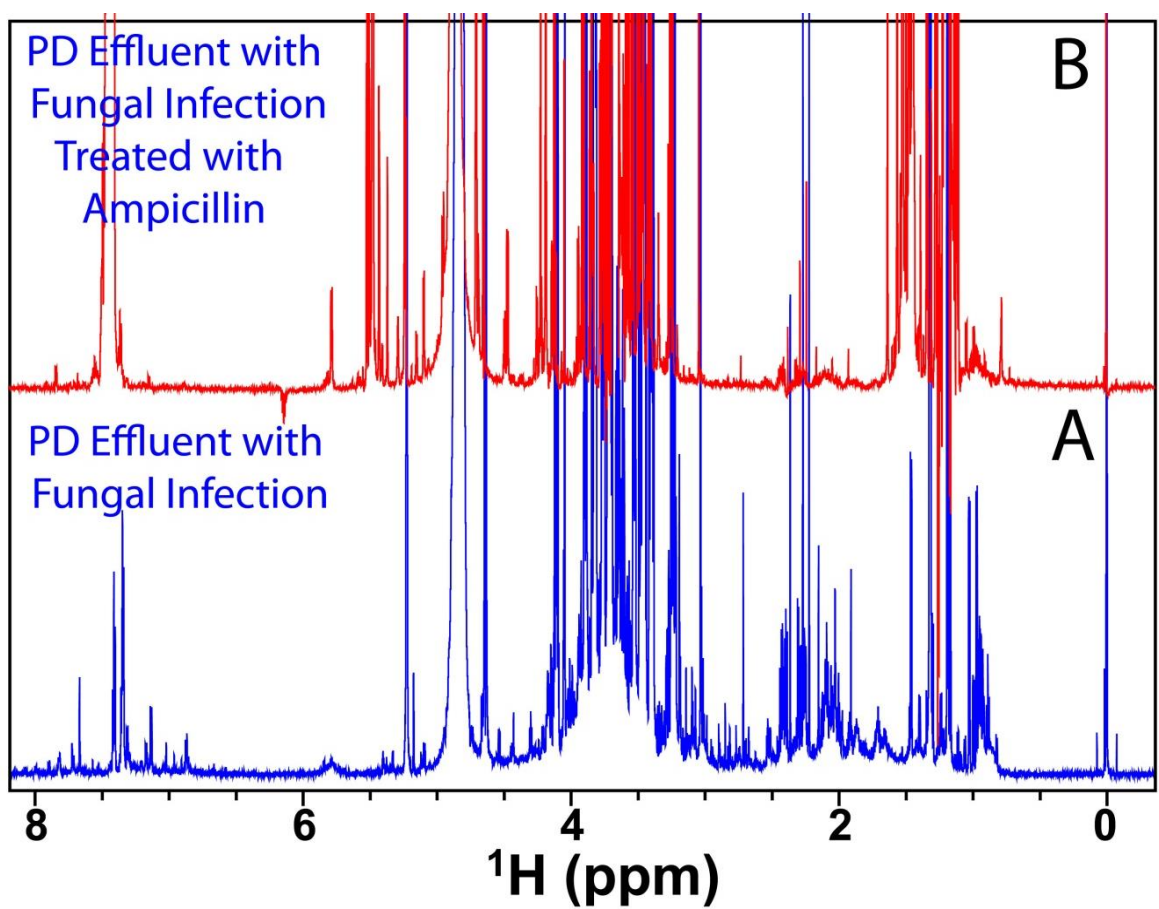

**Figure S5:** Stack plot of one-dimensional (1D) $^1$H CPMG NMR spectra of fungal cell culture: **(A)** without antifungal treatment and sonication, **(B)** supernatant after sonication **(C)** treated with Griseofulvin, and **(D)** treated with Amphotericin.

4. To rule out the possibility that a clinical/biological sample having fungal infection may produce the marker peak when treated with antifungals, *Schizosaccharomyces pombe* (used as a model organism in molecular and cell biology) was cultured (following the routine protocol described here: Sherman, F (2002) Getting started with yeast. Methods in Enzymology 350: 3-41) at 28 °C and 200 rpm in adenine-supplemented YPD (YPAD) media (added with broad spectrum antibacterials: Streptomycin + Streptomycin Sulphate + Penicillin). The fungal cell culture in the exponential phase was transferred to four separate 2 ml centrifuge tubes (1 ml in each tube). One aliquot was directly used to record the 1D $^1$H CPMG NMR spectrum shown in **Fig S6A**, whereas the second aliquot was sonicated and centrifuged at 9000 rpm for 10 minutes, supernatant was used to record the 1D $^1$H CPMG NMR spectrum shown in **Fig S6B**. Third and fourth aliquots were treated with antifungals -named, respectively, Griseofulvin and Amphotericin- at 37 °C for 1 hour and used to record the 1D $^1$H CPMG NMR spectrum shown in **Fig S6C and S6D**. As evident, the fungal cells treated with antifungals does not produce the marker peak **(Fig S6C and S6D)** which thus indicates that the fungal cell membrane lack cyclic fatty acids as stated in this paper.

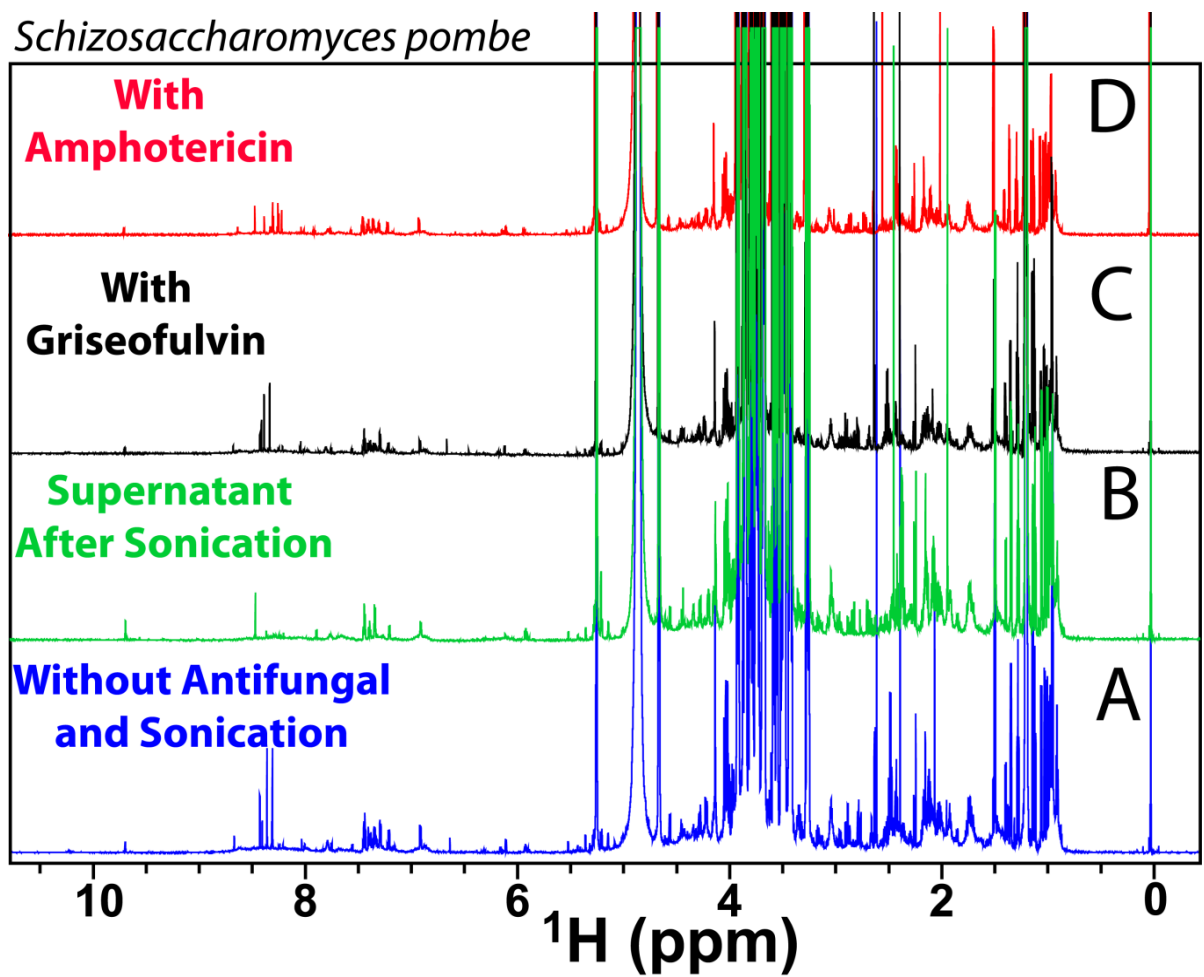

**Figure S6:** Stack plot of one-dimensional (1D) $^1$H CPMG NMR spectra of fungal cell culture: **(A)** without antifungal treatment and sonication, **(B)** supernatant after sonication **(C)** treated with Griseofulvin, and **(D)** treated with Amphotericin.

5. To check if the cyclic fatty acids can be made suspended in the solution using sonication and/or using lysozyme induced cell lysis, the bacterial (*E. coli*) cells in the exponential phase (OD@600nm=0.6) were transferred to four separate 2 ml centrifuge tubes (1 ml in each tube). One aliquot was directly used to record the 1D $^1$H CPMG NMR spectrum shown in **Fig S7A**, whereas the second aliquot was sonicated and centrifuged at 9000 rpm for 10 minutes, supernatant was used to record the 1D $^1$H CPMG NMR spectrum shown in **Fig S7B**. Third aliquot was treated with lysozyme (15,000 units of longlife$^{TM}$ Lysozyme, G Biosciences), incubated at 37 °C for 1 hour and used to record the 1D $^1$H CPMG NMR spectrum shown in **Fig S7C**. Fourth aliquot was also treated with lysozyme (15,000 units of longlife$^{TM}$ Lysozyme, G Biosciences), incubated at 37 °C for 1 hour, sonicated and used to record the 1D $^1$H CPMG NMR spectrum shown in **Fig S7D**. As evident from the spectra **(Fig. S7)**, the marker peak does not appear in any of these conditions suggesting that sonication, lysozyme treatment and even the combination of both these cell lysis treatments, all are inadequate to suspend the membrane fatty acid components into the solution as required here to probe the infection.

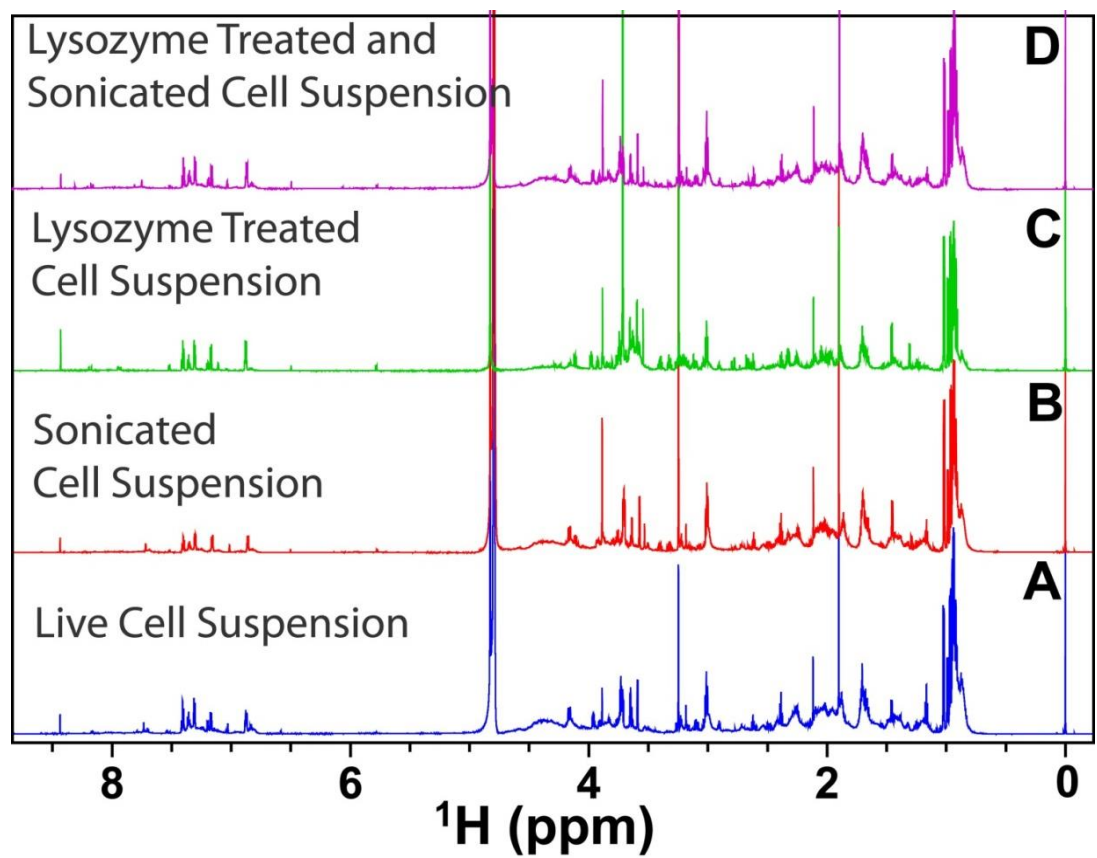

**Figure S7:** Stack plot of one-dimensional (1D) ¹H CPMG NMR spectra of *Escherichia coli* bacterial culture (grown in standard LB media at 37 °C): **(A)** live cell suspension, **(B)** sonicated cell suspension, **(C)** cell suspension treated with lysozyme (at 37 °C for 1 hour) and **(D)** cell suspension treated with lysozyme (at 37 °C for 1 hour) followed by sonication. In each case, the cell suspension obtained in the exponential phase (when OD@600nm = 0.6) has been used.

6. To rule out the possibility that the blood sample when treated with antibiotics may produce the marker peak (because of some kind of biochemical reaction), human blood serum was used to perform another **negative control experiment**. The human serum sample was transferred to two separate 2 ml centrifuge tubes (1 ml in each tube). One aliquot was directly used to record the 1D ¹H CPMG NMR spectrum shown in **Fig S8A**, whereas the second aliquot was treated with ampicillin (antibacterial) for 1 hour at 37 °C and used to record the 1D ¹H CPMG NMR spectrum shown in **Fig S8B**. As evident, the clean/uninfected serum sample treated with ampicillin does not produce the marker peak **(Fig S4B)** further inferring that the marker peak is associated with bacterial/mycobacterial infection.

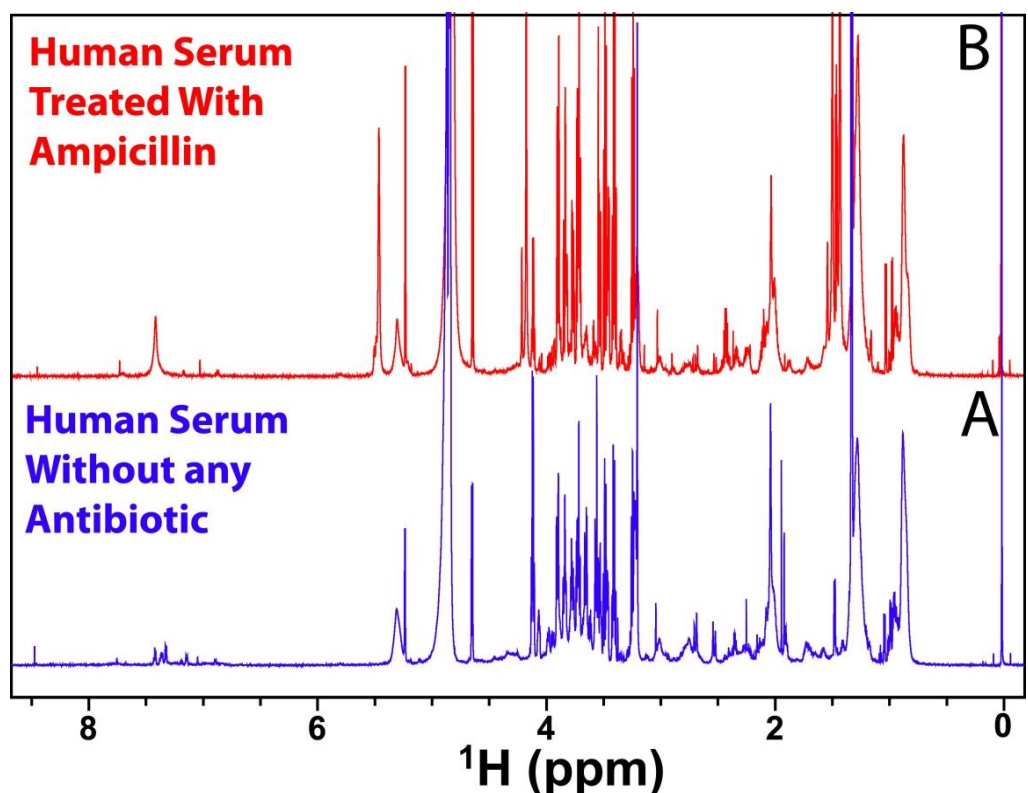

**Figure S8:** Stack plot of one-dimensional (1D) $^1$H CPMG NMR spectra of human serum sample of an healthy: **(A)** without given ampicillin and **(B)** treated with ampicillin (at 37 °C for 1 hour).

7. To check if the proposed **<AADNMR>** method can identify the dead microbial (bacterial/mycobacterial) cells in a sample, sonicated/dead bacterial cell culture was used to perform a test <u>**positive control experiment**</u>. The bacterial (*E. coli*) cell culture in the exponential phase (OD@600nm=0.6) was transferred to a 15 ml centrifuge tube and centrifuged at 8000 rpm. About 13 ml of the supernatant was decanted and the resulted bacterial cell pellet was resuspended in the remaining 2 ml of LB media. The resuspended cells were sonicated and the resulted bacterial cell suspension was transferred to two separate 2 ml centrifuge tubes (1 ml in each tube). One aliquot was directly used to record the 1D $^1$H CPMG NMR spectrum shown in **Fig S9A**, whereas the second aliquot was treated with ampicillin (antibacterial) for 1 hour at 37 °C and used to record the 1D $^1$H CPMG NMR spectrum shown in **Fig S9B**. As evident, the dead bacterial cell suspension treated with ampicillin produce the marker peak **(Fig S9B)** indicating that the proposed method can also be used to identify the dead remains of microbial (bacterial/mycobacterial) infection.

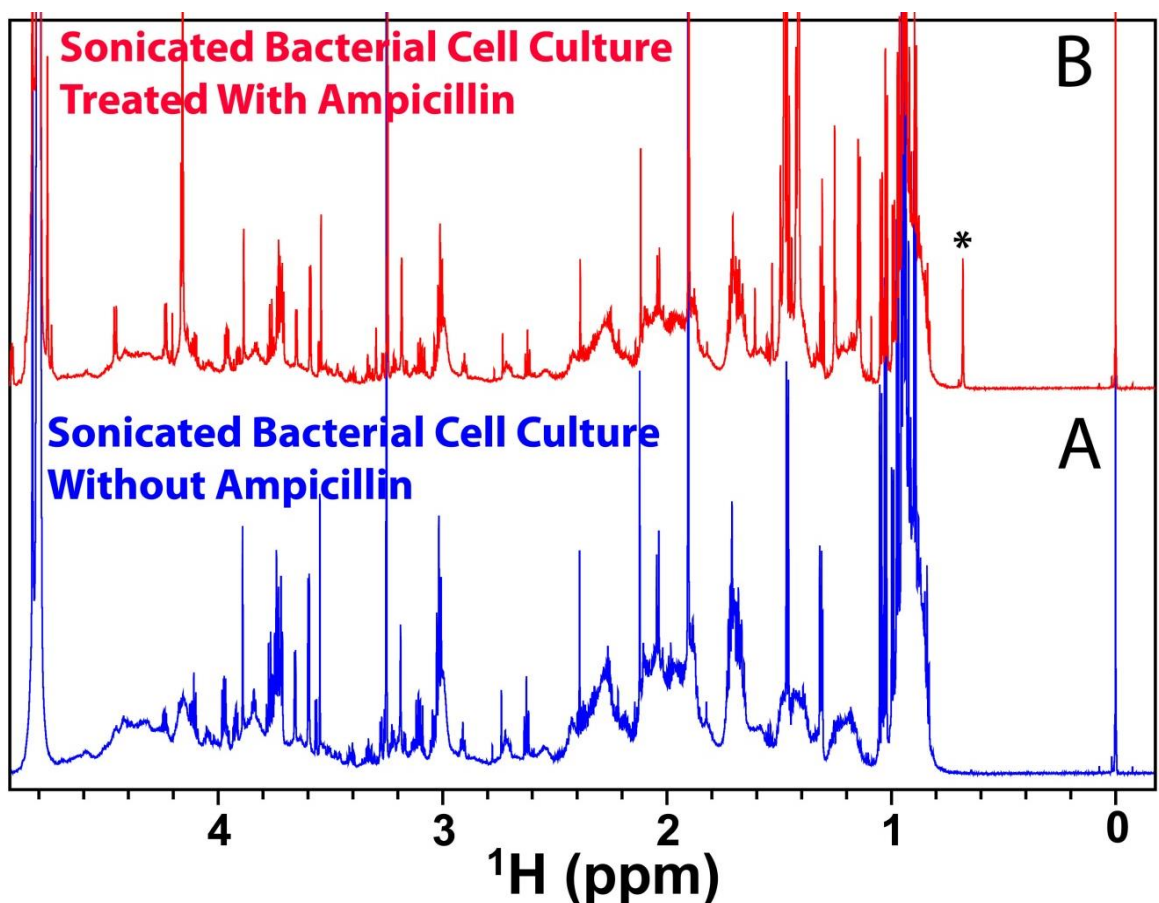

**Figure S9:** Stack plot of one-dimensional (1D) $^1$H CPMG NMR spectra of sonicated *Escherichia coli* bacterial culture (grown in standard LB media at 37 °C: **(A)** without ampicillin treatment and **(B)** treated with ampicillin for 1 hour at 37 °C.

## *Appendix II:*

**Bacterial Cells Cultured in Dianeal Peritoneal Dialysis solution:** In PD effluent obtained from a PD patient having infectious peritonitis and given the intraperitoneal antibiotic treatment, the chemical shift of marker peak was centred about ~0.6 ppm compared to that in the antibiotic treated bacterial cell culture where it is mostly centred about ~0.5 ppm. To check if the intraperitoneal conditions are involved in this chemical shift difference or Dianeal (Baxter U S, containing 2.5 % glucose) PD solution -instilled intraperitoneally in all the PD patients- is causing this chemical shift perturbation, bacterial (*Staphylococcus aureus*) cells were cultured directly into the Dianeal PD solution. Culture in the exponential phase (when OD@600 nm = 0.6) was transferred into two 2 ml centrifuge tubes (1.0 ml culture in each tube). One aliquot was used as a control (representing live cell suspension) and the other aliquot was treated with ampicillin (50 μl of 10 mg/ml stock solution). Both the culture aliquots were again kept in the shaker incubator for about 1 hour under identical conditions (37 °C, 200 rpm). Finally, each culture aliquot was centrifuged for 5 min at 4 °C and 12,000 rpm to remove all cell debris and other contaminants. The supernatant part was decanted and stored at -20 °C until the $^1$H NMR experiments were performed.

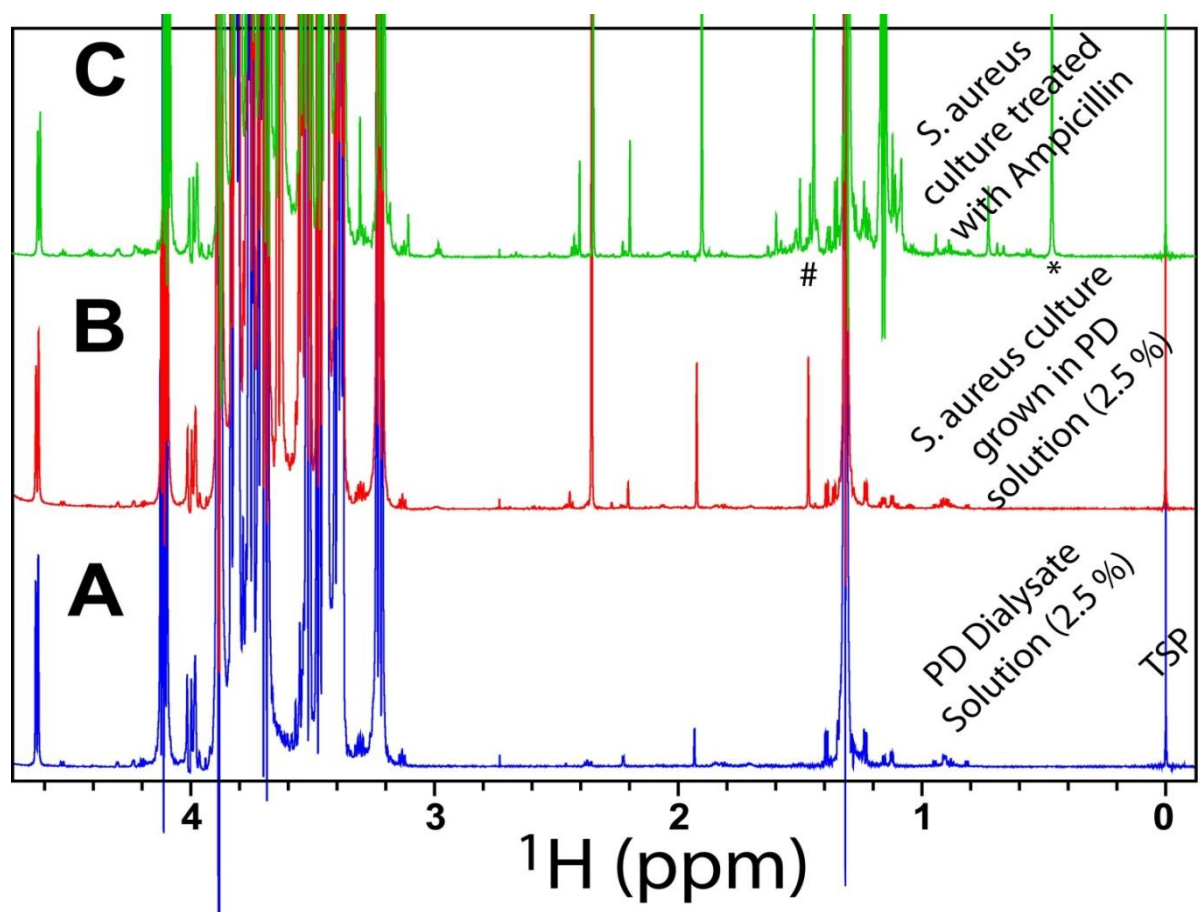

**Figure S10:** Stack plot of one-dimensional (1D) ¹H CPMG NMR spectra of: **(A)** commercial 2.5 % PD dialysate solution and **(B, C)** bacterial cell culture (*Staphylococcus aureus*, grown in the PD solution): **(B)** without antibiotic treatment and **(C)** after incubating the bacterial cell culture with ampicillin at 37 °C for 1 hour. The marker peak in ampicillin treated sample was found to be upfield shifted (close to 0.5 ppm) suggesting that the intraperitoneal conditions are possibly involved in modulating the quality of microbial cyclic fatty acids or the association of ampicillin with fatty acid is causing the upfield shift.